\documentclass[twocolumn,twoside,slac_two]{revtex4}
\usepackage{graphicx}
\usepackage{fancyhdr}

\usepackage{amssymb,amsmath,latexsym,mathrsfs}
\usepackage{xspace}
\usepackage{comment}

\newcommand{\BAR}[1]{\overline{#1}}
\newcommand{\particle}[1]{{\ensuremath{ #1}}\xspace}

\newcommand{\pp}{\particle{pp}}

\newcommand{\bb}{\particle{b\BAR{b}}}

\newcommand{\Bs}{\particle{B^0_s}}
\newcommand{\Bd}{\particle{B^0}}

\newcommand{\Bu}{\particle{B^+}}

\newcommand{\Kst}{\particle{K^{*0}}}  
\newcommand{\Jpsi}{\particle{J\!/\!\psi}}
\newcommand{\Jmm}{\particle{\Jpsi(\mu\mu)}}
\newcommand{\mup}{\particle{\mu^+}}
\newcommand{\mum}{\particle{\mu^-}}

\newcommand{\decay}[2]{\particle{#1\!\to #2}\xspace}

\newcommand{\Bsmumu}{\decay{\Bs}{\mup\mum}}
\newcommand{\BKstmumu}{\decay{\Bd}{\Kst\mup\mum}}

\newcommand{\BdKpi}{\decay{\Bd}{ K^+\pi^-}}

\newcommand{\BuJpsimumuK}{\decay{B^+}{\Jmm K^+}}

\newcommand{\unit}[1]{\ensuremath{\,{\rm #1}}\xspace}

\newcommand{\tev}{\unit{TeV}}

\newcommand{\gevc}{\unit{GeV\!/\!{\it c}}}

\newcommand{\mevcc}{\unit{MeV\!/\!{\it c}^2}}

\newcommand{\invpb}{\unit{pb^{-1}}}
\newcommand{\invfb}{\unit{fb^{-1}}}
\newcommand{\unitlum}{\unit{cm^{-2} \, s^{-1}}}

\newcommand{\fs}{\unit{fs}}
\newcommand{\micron}{\unit{\mu m}}

\newcommand{\BR}[1]{\ensuremath{\mathcal{B}(#1)}\xspace}

\newcommand{\tbe}{\ensuremath{\tan \beta}\xspace}

\begin{document}

\title{Search for New Physics in {\boldmath$B$} Rare Decays at LHCb}

%
\author{Marc-Olivier Bettler, on behalf of the LHCb collaboration}
\affiliation{Ecole Polytechnique F\'ed\'erale de Lausanne, Lausanne, Switzerland}

\begin{abstract}
The LHCb experiment, bolstered up by the $10^{12}$ $b$-hadrons to be produced yearly in its interaction region, is an excellent place to study rare $B$ decays.
Flavor-changing neutral currents are forbidden at tree level in the Standard Model.
They proceed through loop diagrams and hence are indirectly sensitive to New Physics through the effect of new particles on observable quantities.
In this paper, we present preparation studies of the three most promising $B$ rare decay analyzes.
These aim at the observation of  the photon polarization in $B_s \to \phi \gamma$, the measurement of the  angular distribution of the  \BKstmumu decay, and  the search for the yet unobserved $B^0_s \to \mu^+ \mu^-$ decay.
The current analysis strategies and the expected sensitivities are presented.
\end{abstract}

\maketitle

\thispagestyle{fancy}

\section{Introduction}
The Standard Model (SM) is successful in explaining almost all observations in particle physics experiments so far. 
Nevertheless, there are reasons to consider it as a low energy effective limit of a more general theory. 
In that prospect, observables for processes where the SM contribution is highly suppressed are particularly interesting. 
Within the SM, flavor-changing neutral current processes are highly suppressed since they are forbidden at tree level and can only proceed via loops diagrams. 
If New Physics (NP) exists, new particles can also contribute to those processes in the loop diagrams, modifying observable quantities with respect to the SM prediction. 

The LHCb detector~\cite{LHCb} will take advantage of the copious \bb production cross-section expected at LHC~\cite{LHC}. 
It is a single-arm forward spectrometer primarily optimized for the study of $\mathcal{CP}$-violation and rare $b$-hadron decays. 
The detector is characterized by its precise vertex detector, powerful particle identification capabilities and versatile trigger. 
Nominally, LHCb will operate at a luminosity $\mathcal{L} = 2 \times 10^{32}$\unitlum, giving $2$\invfb of data per year ($10^7$ seconds).
The analyzes presented in this document are applied to Monte Carlo simulated data with a full detector response, including pile-up (multiple \pp collisions in a single bunch crossing) and spill-over (signal coming from particles produced in a previous bunch crossing).

$B$ physics requires excellent vertexing capabilities, momentum resolution and particle identification.
The study of rare decays  demands  high background rejection and trigger performance.
LHCb was designed to fulfill these requirements.
Its Vertex Locator provides an impact parameter resolution of $14\micron \pm 35\micron/p_{\rm T}[\gevc]$ that results in a  $B$ proper-time  resolution of $\sigma(\tau) \approx 40$--$100$\fs depending on the decay mode.
The tracking system yields  a momentum resolution of $\sigma(p)/p \approx 0.4\%$, leading to an invariant mass resolution of $20$\mevcc and $15$\mevcc for \Bsmumu and $\Bd \to \Kst\mup\mum$, respectively.

\section{{\boldmath$\Bs \to \phi \gamma$} Photon Polarization}
The branching fraction of this decay, \BR{\decay{\Bs}{\phi \gamma}}$ = \left( 57 ^{+18 \ +12}_{-15 \ -11} \right)\times 10^{-6}$~\cite{Wicht08}, was recently measured by the Belle collaboration and is in agreement with the SM prediction.
In the SM  the polarization of the photon is precisely predicted. 
\decay{\overline{B}^0_s}{X_s \gamma_L} and  \decay{\Bs}{X_s \gamma_R} are allowed transitions under helicity conservation, whereas the crossed transitions are suppressed by a factor $\frac{m_s}{m_b}$.
An observable $\psi$ is defined from the $\overline{B}^0_s$ transition amplitude ratio as
%
%
%
\begin{equation}
\tan \psi \equiv \left| \frac{\mathcal{A}(\decay{\overline{B}^0_s}{\phi  \gamma_{\rm R}})}{\mathcal{A}(\decay{\overline{B}^0_s}{\phi  \gamma_{\rm L}})} \right|, \nonumber 
\end{equation}
which is proportional to $m_s/m_b$ and very small in the SM.
In extensions of the SM, such as the Left-Right Symmetric Model~\cite{phi_gamma_LR} and the unconstrained MSSM~\cite{phi_gamma_uMSSM} predict large values for $\psi$.
The measurement of the photon polarization thus provides a null test for the SM. 

The photon polarization is measured indirectly, through the time-dependent decay rate~\cite{Atwood97}.
For the decay of \Bs ($\overline{B}^0_s$) mesons, the decay rate is:
\begin{eqnarray}\label{equ:phigamma}
\Gamma_{\{ \Bs,\overline{B}^0_s \} \to \phi \gamma} (t) & \propto & e^{- \Gamma_s t} \Big\{ \cosh \frac{\Delta\Gamma_s t}{2} - \mathcal{A}^\Delta \sinh \frac{\Delta\Gamma_s t}{2}  \nonumber \\
& & \pm \mathcal{C} \cos \Delta m_s t \mp \mathcal{S} \sin \Delta m_s t\Big\}.
\end{eqnarray}
Unlike Belle and {\sc BaBar} analyzes studying $\mathcal{CP}$ asymmetry in $\Bd \to K^0_S \pi^0 \gamma$~\cite{Ushiroda06,Aubert08} decay, we consider an analysis without flavor tagging.
Thus, adding the  \Bs and $\overline{B}^0_s$ contributions, the two last terms of Eq.~(\ref{equ:phigamma}) cancel out.
Furthermore, since in the SM $\Delta\Gamma_s$ is expected to be non-zero in the \Bs system, the decay rate is sensitive to $\psi$, through $\mathcal{A}^\Delta = \sin 2 \psi \cos \varphi$~\cite{Muheim08}. 
Here $\varphi$ represents the decay weak phase, that is approximately zero in the SM. 
Note that, in the SM, a flavor tagged analysis is not expected to provide a better sensitivity on the measure of $\psi$, because $\mathcal{C} < 1\,\%$ and $\varphi \approx 0$.

The main experimental issue is to understand the bias on the \Bs lifetime coming from the trigger and offline cuts for the $\phi$ selection.
For 2\invfb, the expected signal yield is 11000 events with a $B/S < 0.6$ and a \Bs mass resolution of 92\mevcc~\cite{phi_gamma_note1}.
This yield, leads to an expected resolution of 0.1 for the photon polarization observable $\psi$~\cite{phi_gamma_note2}.

There are others channels considered to probe photon polarization, such as \decay{\Lambda_b}{\Lambda^0 \gamma}, $\Lambda_b \to \Lambda^{\ast}( \to pK^-) \gamma$, and  \decay{B^+}{\phi K^+ \gamma}.


\section{{\boldmath$\BKstmumu$} Angular Analysis}

This decay branching fraction \BR{\BKstmumu}$ = \left( 9.8 \pm 2.1 \right) \times 10^{-7}$~\cite{PDGlive} is in agreement with the SM.
Asymmetries related to angular distributions are theoretically well predicted because hadronic uncertainties cancel out in the ratios.
The forward-backward asymmetry of $\theta_{\ell}$, the angle between the $\mu^-$ and the $\overline{B}^0$ flight direction in the di-muon rest frame is such an observable.
Thus, the shape of $A_{\rm FB}$ as a function of the di-muon invariant mass squared, $q^2$, is calculated for the SM and for extension models. 
The zero-crossing point of $A_{\rm FB}$ ($q_0^2$ such that $A_{\rm FB}(q_0^2) = 0$) is a particularly interesting observable. 
Its SM prediction  is $q^2_0 = 4.36 ^{+0.33}_{-0.31}$\unit{GeV^2\!/\!{\it c}^4}~\cite{qzero}.

The $A_{\rm FB}$ for this channel was already measured at {\sc BaBar}, Belle and CDF~\cite{Babar_Kmumu, Belle_Kmumu, CDF_Kmumu} and exhibits some tension with respect to the SM prediction.
However, the statistics is very low: the Belle analysis, the most sensitive so far, is based on 230 events only.  
LHCb expects 7000 events per 2\invfb, with $B/S \approx 0.2$.

The challenge for this analysis is the understanding of the biases on angular observables induced by detector and reconstruction effects.
It is worth to note that the zero-crossing point value is little sensitive to those biases, making its measurement particularly interesting with early data.

\begin{figure}[t]
\includegraphics[width=\linewidth]{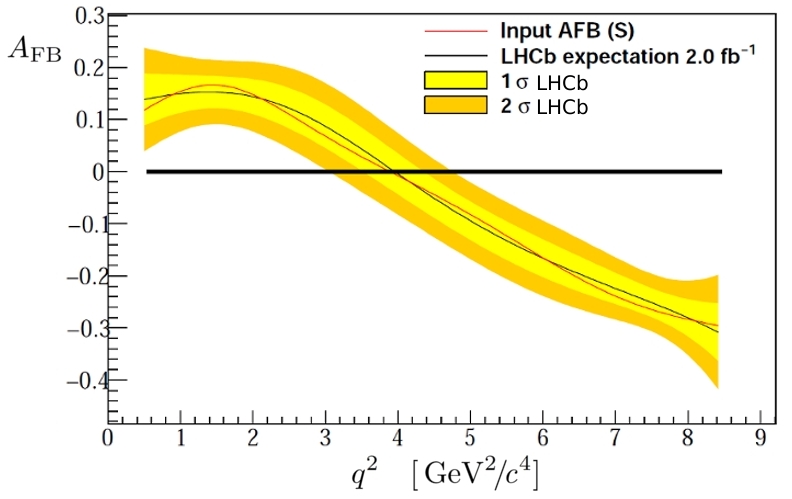}
\caption{\label{Kmumu_unbinned}LHCb sensitivity to $A_{\rm FB}$ with the unbinned \BKstmumu analysis for 2\invfb, with $1\sigma$ and $2\sigma$ LHCb experimental error bands, allowing the extraction of the zero-crossing point value $q^2_0$.
This simulation assumes the SM with $q^2_0 = 3.97$\unit{GeV^2\!/\!{\it c}^4}.
}
\end{figure}

The LHCb strategy for this decay consists of different methods of increasing sensitivity but also increasing requirements in statistics and acceptance understanding.

The first method is a binned analysis for $A_{\rm FB}$, combined with a linear fit to extract the zero-crossing point  value~\cite{Kstar_note1}.
With a few hundreds of \invpb, this method will already compete with current measurements. 
With 2\invfb, the expected resolution on $q^2_0$ is $0.5$\unit{GeV^2\!/\!{\it c}^4} and LHCb expects to reach the theoretical precision with 10\invfb of data.

The second method consists of an unbinned analysis for $A_{\rm FB}$, with no linear assumption around the zero-crossing point~\cite{Kstar_note2}.
This analysis yields a  sensitivity similar to the binned analysis.
Figure~\ref{Kmumu_unbinned} depicts the expected $A_{\rm FB}$ shape obtained by the unbinned analysis for the SM and with 2\invfb.

The next step is a full angular analysis, based on the three angles, which, with the di-muon invariant mass, completely define the decay kinematics~\cite{Kstar_note3}.
Using the three angle distributions, a number of asymmetries and transversity amplitudes can be computed.
The full angular analysis requires a good knowledge of detector acceptance effects and an integrated luminosity of at least 2\invfb.

\begin{figure}[t]
\includegraphics[width=\linewidth]{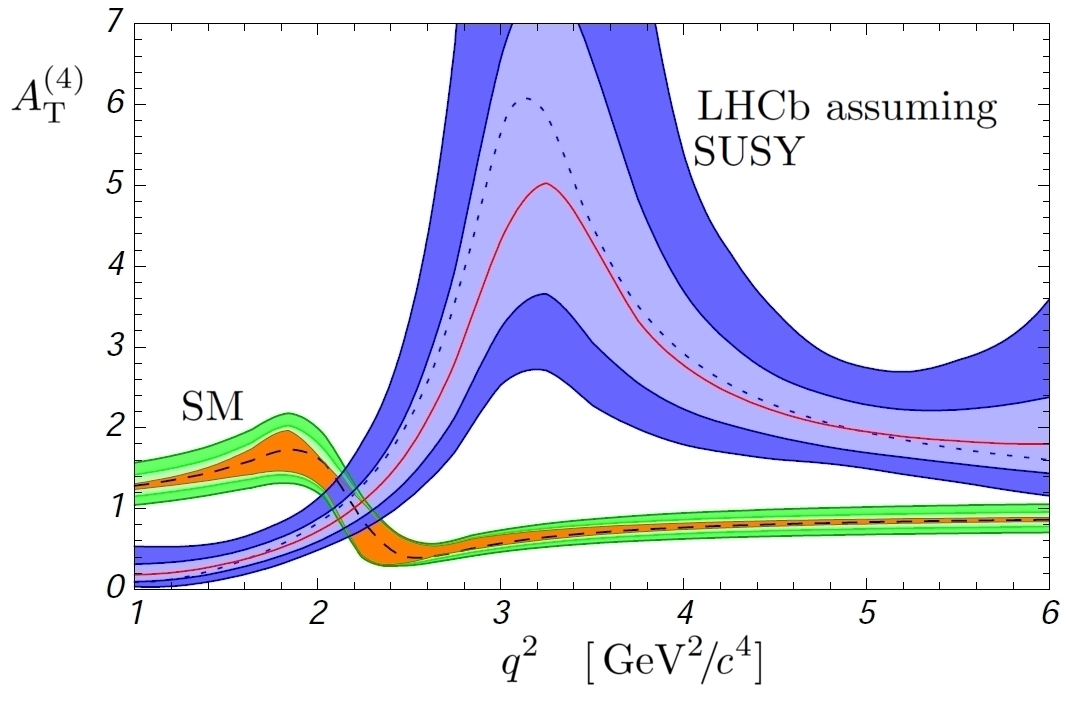}
\caption{\label{Kmumu_AT4}$A^{(4)}_{\rm T}$ observable, defined in~\cite{Egede_theo}.
The SM prediction is shown by the dashed line, with $1\sigma$ and $2\sigma$ theoretical uncertainty bands.
The dotted line depicts a particular SUSY scenario with positive mass insertion and large gluino mass, with $1\sigma$ and $2\sigma$ LHCb experimental error bands (the solid line showing the central value of the toy experiments).
}
\end{figure}

A thorough study has assessed the sensitivity to various observables~\cite{Egede_theo}, of which some were shown to differ significantly between SM and NP scenarios.
An example of the sensitivity achievable with 10\invfb for an  observable particularly sensitive to NP is shown in Fig.~\ref{Kmumu_AT4}.

\section{{\boldmath$\Bsmumu$} Branching Fraction}
Within the SM, the \Bsmumu  branching ratio is expected to be $\BR{\Bsmumu} = ( 3.35 \pm 0.32)\times 10^{-9}$~\cite{Blanke06}. 
In the MSSM, this decay receives additional contributions from new particles, {\it e.g.} for large \tbe values, the ratio of the Higgs vacuum expectation values, the branching ratio is then proportional to \tbe to the sixth power, and can be considerably enhanced~\cite{Ellis07_WMAPMSSM}.
The latest upper limit given by the CDF experiment is $3.6 \times 10^{-8}$ at $90\%$ CL~\cite{mumu_CDF}.

\begin{figure*}[t]
\begin{minipage}[b]{0.47\linewidth}
\centering
\includegraphics[width=\linewidth]{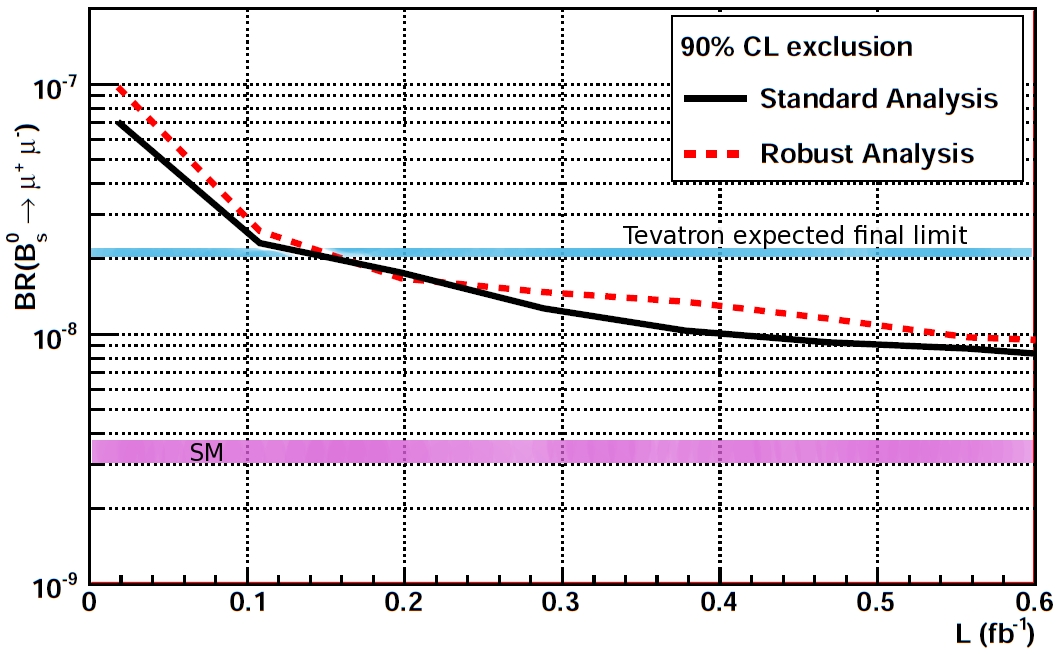}
\end{minipage}
\begin{minipage}[b]{0.51\linewidth}
\centering
\includegraphics[width=\linewidth]{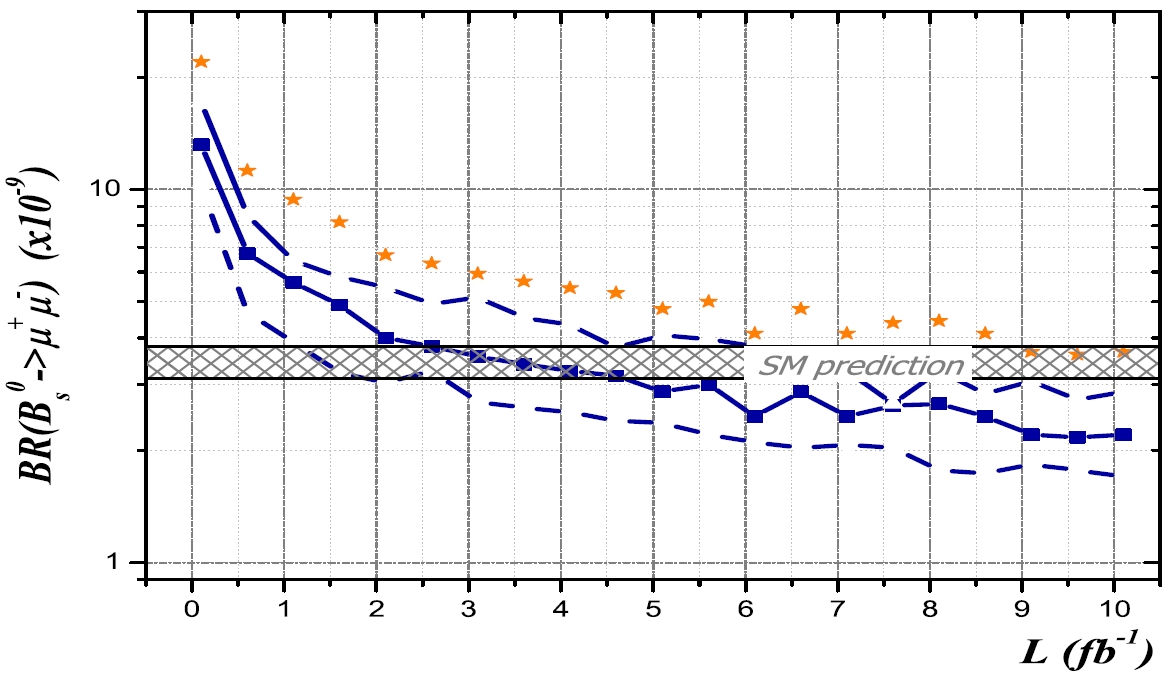}
\end{minipage}
\caption{\label{sensitivity}Expected $90\%$ CL upper limit on \BR{\Bsmumu} in absence of signal for 8\tev collisions (left) and \BR{\Bsmumu} at which a $5\sigma$ discovery (stars) or a $3\sigma$ evidence (solid blue curve) is expected for 14\tev collisions (right), as a function of the integrated luminosity.
On the exclusion plot, the black curve is the result of the standard analysis, and the red dashed curve is the result of the robust analysis.
The background estimate has conservatively been set to its  $90\%$ CL upper limit in the exclusion plot, and similarly the dashed curves indicates the 90\% CL upper and lower limit in the observation case. 
The Tevatron limit is calculated by extrapolating the current result to 8\invfb per experiment. 
}
\end{figure*}

The current LHCb strategy for the \Bsmumu decay search~\cite{mumu-1} can be summarized as follows.
An efficient selection is applied with the goal to remove obvious background with minimal loss of signal events.
The analysis relies on three independent variables related to,  the di-muon invariant mass, the particle identification of the daughters, and geometrical information from the decay topology. 
Since the three variables are uncorrelated, they can be calibrated independently. 
For each event, likelihood values are calculated for each of the three variables, under the signal and background hypotheses.
Calibration methods for these values, both for signal and for background, are designed to rely solely on real data.
The compatibility of the obtained likelihood distributions is tested against various \Bsmumu branching ratio hypotheses, using the \emph{CL} modified frequentist analysis~\cite{Cls}. 
The final result is either a measurement or an upper limit of the branching fraction.

The branching fraction is expressed as:
\begin{equation}
\BR{\Bsmumu} = \frac{ \mathcal{N}_{\rm sig} }{ 2 \ \sigma_{b\overline{b}} \ \mathcal{L}_{\rm int} \ f_s \  \epsilon_{ \rm sig } }, \nonumber
\end{equation}
where  $\mathcal{N}_{\rm sig}$ is the number of signal events, $\sigma_{b\overline{b}}$ is the $b\overline{b}$ production cross-section, $\mathcal{L}_{\rm int}$ the integrated luminosity,  $ f_s$ the probability of a $b$ quark to hadronize in a \Bs meson, and $\epsilon_{\rm sig}$ the product of the reconstruction, trigger, and selection efficiencies. 
Since the number of \Bs produced, $2\sigma_{b\overline{b}} \ \mathcal{L}_{\rm int}$, will not be precisely known, the use of a normalization channel of well-known branching fraction is required to obtain an absolute measurement of the branching fraction or upper limit.

The \mbox{\Bsmumu} branching fraction is then:
\begin{equation}
\BR{\Bsmumu} =\mathcal{B}_{\rm norm} \times \frac{f_{\rm norm}}{f_s} \times  \frac{\epsilon_{\rm norm}}{\epsilon_{\rm sig}} \times \frac{ \mathcal{N}_{\rm sig} }{\mathcal{N}_{\rm norm}} , \nonumber
\end{equation}
where $f_{\rm norm}$, $\epsilon_{\rm norm}$ and $\mathcal{N}_{\rm norm}$ are the quantities for the normalization channel, with definitions analogous to those of the signal channel. 

Possible normalization channels are \BuJpsimumuK and \BdKpi.
Particular care has been taken to analyze the signal and normalization channels in a common way such as to cancel any large systematic effects in the efficiency ratio. 

The main systematics arise from the $\sim 13 \%$ uncertainty on the ratio $\frac{f_{\Bd}}{f_s}$ or $\frac{f_{\Bu}}{f_s}$. 
With sufficient statistics, all other systematic uncertainties are expected to get much smaller than that, as the analysis relies solely on data.
To eliminate the factor $\frac{f_{\rm norm}}{f_s}$, the \decay{\Bs}{D^-_s \pi^+} decay mode could be used in the future a normalization channel, if its absolute branching fraction can be measured more precisely at Belle.

The expected sensitivity of LHCb to the \Bsmumu decay as a function of the integrated luminosity is shown in Fig.~\ref{sensitivity}.
The solid line in the left plot shows the expected upper limit, at 90\% confidence level, on the branching fraction when no signal is observed, for $pp$ collisions at $\sqrt{s}= 8$\tev.
The branching fraction for which a $5\sigma$ discovery (or for which  a $3\sigma$ evidence) is expected, is shown on the right plot for 14\tev collisions.

The left plot also indicates the expected limit from the Tevatron experiments, extrapolating the current results to 8\invfb of data per experiment.
It shows that LHCb can compete with the Tevatron with approximately  0.1\invfb of data and overtake its expected limit with about 0.2\invfb. 
In that time frame, the detector may not be fully understood yet. 
Therefore, an alternative robust analysis is designed using variables with a  physical content similar to the ones of the standard analysis, but avoiding the use of error estimates. 
This implies a modified selection and definition of the geometrical likelihood.
The robust analysis sensitivity is also depicted in Fig.~\ref{sensitivity} (left), as a red dashed curve.
The robust analysis presents  a sensitivity comparable with the one of the standard analysis. 
Therefore, it constitutes a valuable option for the early data. 

About 3\invfb (10\invfb) are enough for a $3\sigma$ evidence ($5\sigma$ observation) if the branching fraction is equal to the SM prediction.
Any enhancement driven by NP will be observed sooner.
Particularly, if the branching fraction is as high as $2\times10^{-8}$, as predicted in Ref.~\cite{Ellis07_WMAPMSSM}, a $5\sigma$ discovery is possible with very little luminosity ($< 0.4$\invfb).

\section{Conclusion}
The large \bb production cross-section expected at LHC, together with the characteristics of LHCb, leads to a good performance in the search for NP in FCNC decays.
Rare $B$ decays are particularly interesting because of the precise theoretical predictions for some of their observables, and they provide stringent test of the SM and its extensions.  
Expected sensitivities for LHCb's analyzes of the most promising rare $B$ decays were presented.

One year of data taken in nominal conditions yields to the observation of the photon polarization in the \decay{\Bs}{\phi \gamma} decay to a 10\% level and to the measurement of the $A_{\rm FB}$ zero-crossing point to a $0.5$\unit{GeV^2\!/\!{\it c}^4} resolution in the \BKstmumu decay mode.
The same amount of data allows to set a limit on the \Bsmumu branching fraction down to the SM prediction if no signal is observed, strongly constraining NP models with high \tbe values.
In case the branching fraction is enhanced, an observation of NP is possible.

In the long term, a full angular analysis of the \BKstmumu can be performed and bringing enhanced sensitivity in probing for NP.
A $5\sigma$ discovery of the \Bsmumu decay is possible if the branching fraction is at the SM level with about 10\invfb.
Any NP enhancement will be discovered before.

\bigskip 

\end{document}